\documentclass[runningheads]{llncs}
\usepackage{graphicx}
\begin{document}
\title{Social Network Analysis of the Professional Community Interaction - Movie Industry Case}
\titlerunning{Network Analysis of the Movie Industry Community}
\author{Ilia Karpov \inst{1}\orcidID{0000-0002-8106-9426} \and
Roman Marakulin \inst{1}\orcidID{0000-0003-0738-1267}}
\authorrunning{Ilia Karpov and Roman Marakulin}
\institute{Higher School of Economics, Moscow, Russia
\email{\{karpovilia,marakromrom\}@gmail.com}}

\maketitle              % typeset the header of the contribution
\begin{abstract}
With the rise of the competition in the movie production market, because of new players such as Netflix, Hulu, HBO Max, and Amazon Prime, whose primary goal is producing a large amount of exclusive content in order to gain a  competitive advantage, it is extremely important to minimize the number of unsuccessful titles. This paper focuses on new approaches to predict film success, based on the movie industry community structure, and highlights the role of the casting director in movie success. Based on publicly available data we create an ``actor''-``casting director''-``talent agent'' - ``director'' communication graph and show that usage of additional knowledge leads to better movie rating prediction.

\end{abstract}
\section{Introduction}
The movie industry is undergoing significant changes. Online streaming platforms such as Netflix, HBO Max, Hulu, Disney Plus, Amazon Prime and others, have become extremely popular. The main goal of these platforms is to get as many recurrent users as possible by creating exclusive titles that are only available on one of the platforms, and the amount of content will increase rapidly over the next few years and competition between movie producers will become more fierce.

In order to produce greater amounts of movie titles, producers must invest more in production, considering that it is an extremely risky and expensive type of investment. Media companies or streaming platforms are eager to minimize the number of unsuccessful titles because they will lose their competitive advantage, and lose their market share. This leads to new models and approaches to the estimation of future success of a movie based on the data that is available before a company invests in a project

Traditional solutions classify a future title’s success based on statistical information like budget, genre, duration, and so on. The movie industry seems to be quite a close community where a lot of people interact with each other by taking part in similar projects or having the same talent agents – the people who find jobs for an actor and process the incoming offers. To the best of our knowledge, there are no specific approaches to predict movie success when taking into account the information about title principals and their position in the movie industry community.  Those features seem to be important because it may be useful to have a casting director who has authority in the movie community and will more likely be able to gather the best cast that will lead a movie to success. However, success may be considered in different ways: academy award, earnings, and rating. In this study, success will be measured by movie ratings taken from IMDb, because awards are often given to movies that meet the current public agenda. Earning strongly depends on marketing spending and brand popularity, for example “The Avengers: End Game” had a \$200 million marketing budget with an extremely popular brand and 220 million production budget that led to \$1.6 billion in box office revenue.

The main contributions of the paper are the following:
\begin{itemize}
    \item We propose a joint dataset that incorporates statistical and social network data from three different sources. To the best of our knowledge, this is the first dataset that takes into account social contacts of casting directors and talent agents.
    \item We study the network structure of the movie community and use the obtained information to predict the future success of a movie title. Our experiment shows that social network information can improve classification accuracy from 4 to 6 \% depending on the classification method
    \item We find out that a casting director is more important than actors in terms of feature importance for the film rating. This leads to the opportunity of predicting the movie’s success at an early stage when actors are not even approved for the role.
\end{itemize}

The rest of this paper proceeds as follows. Section 2 contains an overview of the previous studies regarding movie community analysis and a titles’ rating prediction. Section 3 describes the obtained dataset. The 4-th Section contains the details of the movie community graph generation and description. Section 5 is devoted to a movie success classification model, and the conclusion is expressed in  6 Section.

\section{Related work}
In this chapter, previous studies related to graph-based movies analysis and its success prediction are observed.

Michael T. Lash and Kang Zhao \cite{Lash2016EarlyPO} were focused on feature engineering that allowed them to predict the success of a movie more accurately. They built a model that classified movies into 2 groups depending on ROI. As a result, they have achieved a ROC-AUC 0.8-0.9. The data sample contained 14,097 films with 4,420 actors. The independent variables were mainly monetary, for example, the average profit of films for the period, how much the actors earn on average from a film, etc. The social network analysis was used to extract indications of the interactions between actors and directors and use it as features in a classification model.

The authors of the paper “We Don’t Need Another Hero - Implications from Network Structure and Resource Commitment for Movie Performance”, have made a similar work \cite{Meiseberg2007WeDN}. They used common movie attributes such as year, budget, and etc. Also, they used SNA methods to extract features about the interactions between actors in movies, for example, the normalized number of contacts of a given film’s crew with teams from other movies.

Krushikanth R. Apala and others \cite{Apala2013PredictionOM} collected data from social networks, such as comments on trailers on YouTube, the popularity of actors on their Twitter pages and so on, and predicted the success of the film in terms of money. As the main result, they showed that the popularity of the actors is a very important factor in the success of a film.

Bristi, Warda Ruheen, Zakia Zaman, and Nishat Sultana \cite{Bristi2019PredictingIR} have been trying to improve IMDb rating prediction models using five different machine learning approaches: bagging and random forest classifiers, decision tree classifier, KNearest Neighbours and Naïve Bayes classifiers. Also, the researchers divided movie titles into classes based on their IMDb rating (Figure 5), making it a dependent variable for classifiers. In order to handle class imbalance, authors used the Synthetic Minority Oversampling Technique (SMOTE) algorithm which balances classes by taking random observations from datasets that are close in feature space, and generating new samples by linear interpolation between selected data records. Also, the researchers used resampling. As a result, they achieved a 99.23 accuracy metric using random forest. However, there were only 274 movie titles in the dataset and the results were provided for a training dataset only.

Another attempt to predict IMDb scores was taken by Rudy Aditya Abarja and Antoni Wibowo in their article “Movie Rating Prediction using Convolutional Neural Network based on Historical Values” \cite{Abarja2020}. The main idea of the article is to implement Convolutional Neural Network (CNN) to predict a movie’s IMDb rating. Authors utilized the IMDb movie data from Kaggle which contained historical features, such as average movies rating by director, an actor’s average movie rating, average genre rating and so on. Also, some metadata features, such as budget, duration and release date, were used. As a result, the researchers achieved 0.83 MAE, which is 0.11 better than the best baseline model with 0.94 MAE. Also, Ning, X., Yac, L., Wang, X., Benatallah, B., Dong, M., \& Zhang, S showed \cite{NING202012} that deep learning models outperform classical machine learning baselines on the IMDb datasets.

\section{Dataset Exploration}
In this paper we use joint features from several publicly available databases. The first part of the dataset was obtained from the official IMDb web page which provides the following information: title name, language, production year, duration in minutes, genre, director name, writer name, film crew, including persons primary profession, name, and date of birth, IMDb rating, and number of votes.  The second part of the dataset is from the Rotten Tomatoes database that adds information about a movie statistic, such as tomatometer rating, number of votes, number of positive votes, and number of negative votes. This information is used as the retrospective data about a film crew’s previous projects. The third part of the dataset is IMDb Pro – an extension to IMDb that contains more complete information about people in the movie industry, for example, talent agent contacts, filmography with current status for every movie, his earning from a movie title, extended biography, and some ratings such a STARmeter and number of news articles, which shows how popular an actor is. The data about his talent agent’s contact allows to make a connection between an actor and a talent agent. 
The combined dataset contains information about 85,855 movie titles.

Based on this data the following features were extracted:
\begin{itemize}
    \item Country in which a movie has been produced;
    \item Genre of a movie title, for example, horror or drama;
    \item Title, that is the name of the title and its IMDb ID;
    \item Type of production, for example, movie or series;
    \item Year a movie title has been produced;
    \item Duration, that is a movie title runtime in minutes;
    \item Primary language of the movie/series;
    \item Movie crew, that is actors’ names and their IMDb IDs, director name and IMDb ID
    \item IMDb user rating;
    \item Number of votes;
    \item Metascore user rating;
    \item Number of reviews from critics and users on IMDb;
    \item Budget of a movie that has been spent on production;
    \item Income, that is gross income worldwide;
    \item User rating on RottenTomatoes;
    \item Age rating (PG, PG-13 and etc.)
\end{itemize}

Most of the budget, USA gross income, metascore, and worldwide gross income values are missing. 

\begin{figure*}
\centering
\begin{minipage}{.5\textwidth}
  \centering
  \includegraphics[width=1.0\linewidth]{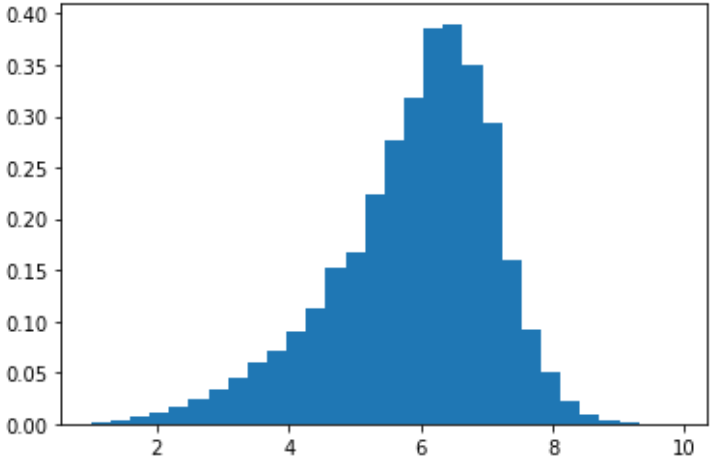}
  \caption{IMDb rating distribution}
  \label{fig_1}
\end{minipage}%
\begin{minipage}{.5\textwidth}
  \centering
  \includegraphics[width=1.0\linewidth]{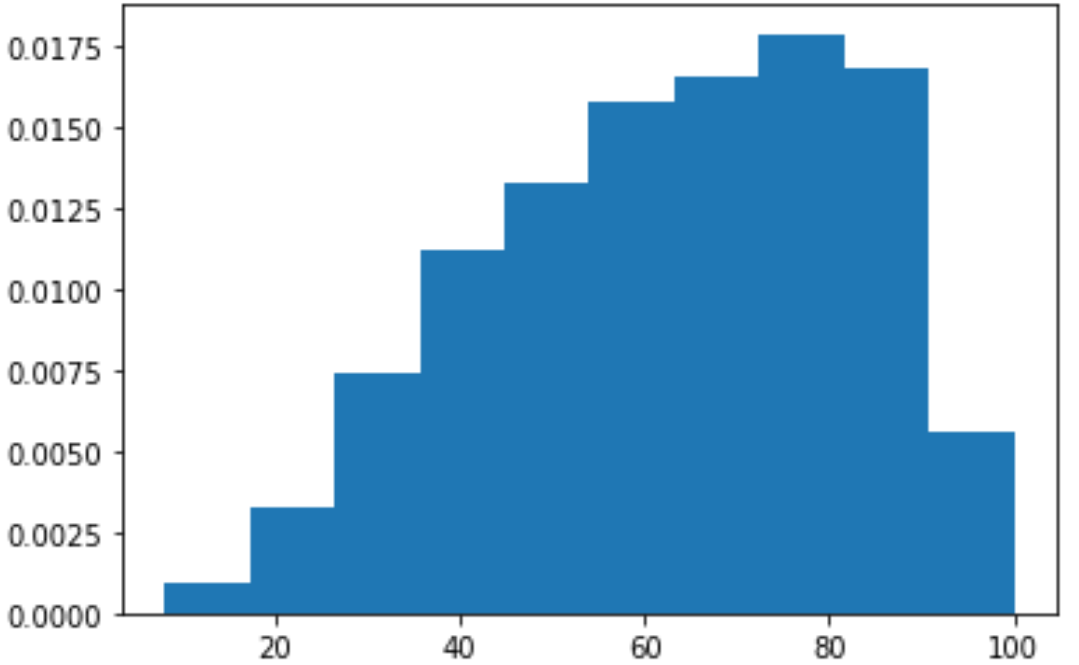}
  \caption{RottenTomatoes user rating distribution}
  \label{fig_1_rt}
\end{minipage}
\end{figure*}

IMDb score distribution is left skewed but quite close to normal \ref{fig_1} There is no need to try to make the distribution normal, because in this research the classification problem is considered, so the IMDb scores will be divided into the following four intervals: 0-3, 4-5, 6-7, 8-10. Rotten Tomatoes users more often rate films with a lower number of points (Figure \ref{fig_1_rt}), making the distribution quite close to the IMDb rating distribution.

\begin{figure*}
\centering
\begin{minipage}{.5\textwidth}
  \centering
  \includegraphics[width=1.0\linewidth]{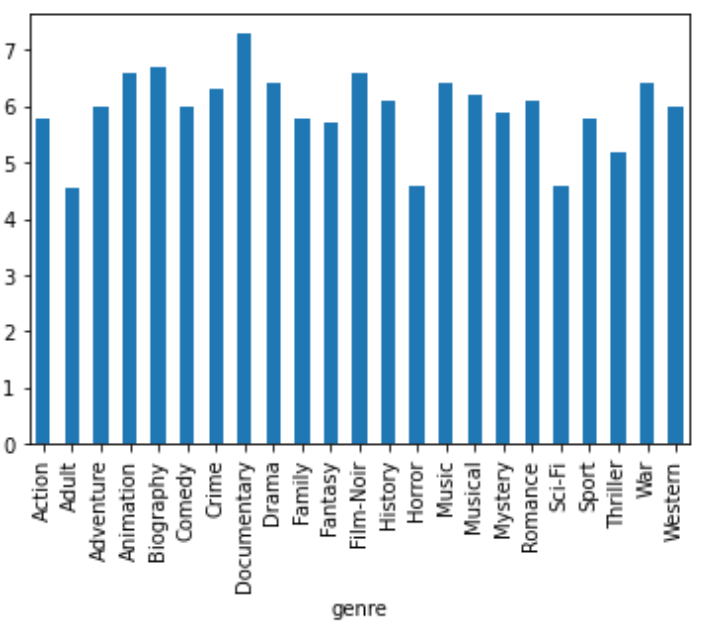}
  \caption{IMDb rating among genres}
  \label{fig_2}
\end{minipage}%
\begin{minipage}{.5\textwidth}
  \centering
  \includegraphics[width=1.0\linewidth]{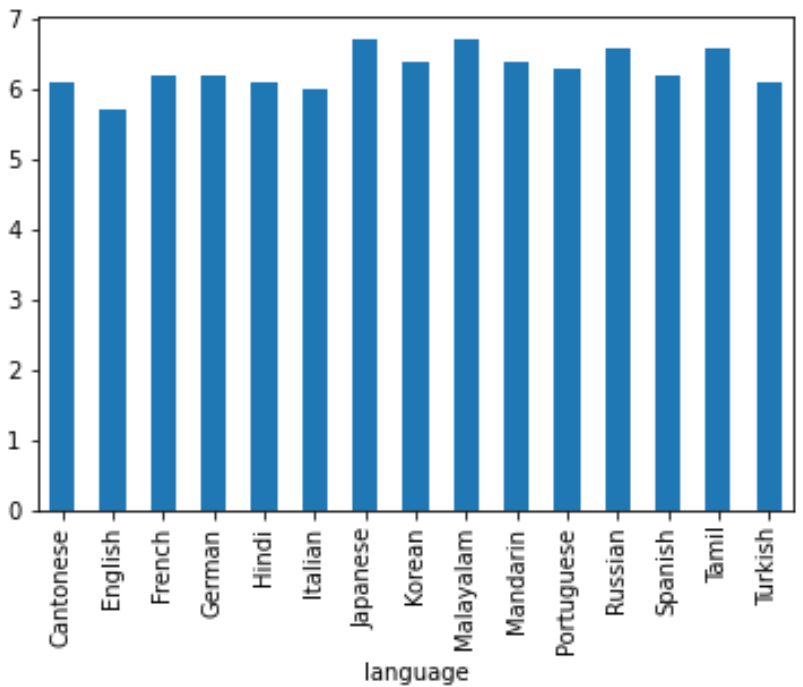}
  \caption{IMDb rating among languages}
  \label{fig_3}
\end{minipage}
\end{figure*}

As can be seen in Figure \ref{fig_2}, the IMDb score does not heavily depend on genre. Only adult, horror, and sci-fi movies have significantly lower average IMDb scores, however, these genres make a small portion, around 1\%, of all records. Also, it can be noted that documentary movies have a better average IMDb rating, which is probably because these movies are usually filmed specifically for smaller groups of people, for example, the National Geographic fans, and they are not often presented in cinemas or featured on Netflix and other streaming platforms. Because of this, they hold a small portion of observations in the dataset.

Language also does not have a significant influence on the IMDb score (Figure \ref{fig_3}), which is quite obvious and leads to the intuition that it does not matter where a movie was produced, and if it is good, it gets high rating, if it is bad, it gets low rating. Most of the movie titles have English, French, Spanish and Japanese as primary languages.

\section{Network Model}

\subsection{Graph Generation}

In order to analyze the movie industry community, it should be presented in the form of a directed graph by the following rules:
\begin{itemize}
    \item Actors, directors, casting directors and talent agents are nodes;
    \item Actor – actor link appears when two actors mutually take part in a movie title, the link is bidirectional;
    \item Actor – director link appears when an actor and director mutually take part in a movie title as actor and director respectively,. The direction of the link comes from director to actor, because a director approves an actor for the role in a movie;
    \item Actor – talent agent link appears when an actor is the client of a talent agent,. The direction is from actor to talent agent, because agents work for actors, assigning them to roles and castings;
    \item Actor – casting director link appears when an actor and casting director mutually take part in a title, as actor and casting director respectively. The direction is from casting director to actor because casting directors find actors who would fit a role;
    \item Director – casting director link appears when director and casting director mutually take part in a movie title as director and as casting director respectively. The direction is from director to casting director because director approval is required.
\end{itemize}

\begin{figure*}
\centering
\includegraphics[width=1.0\linewidth]{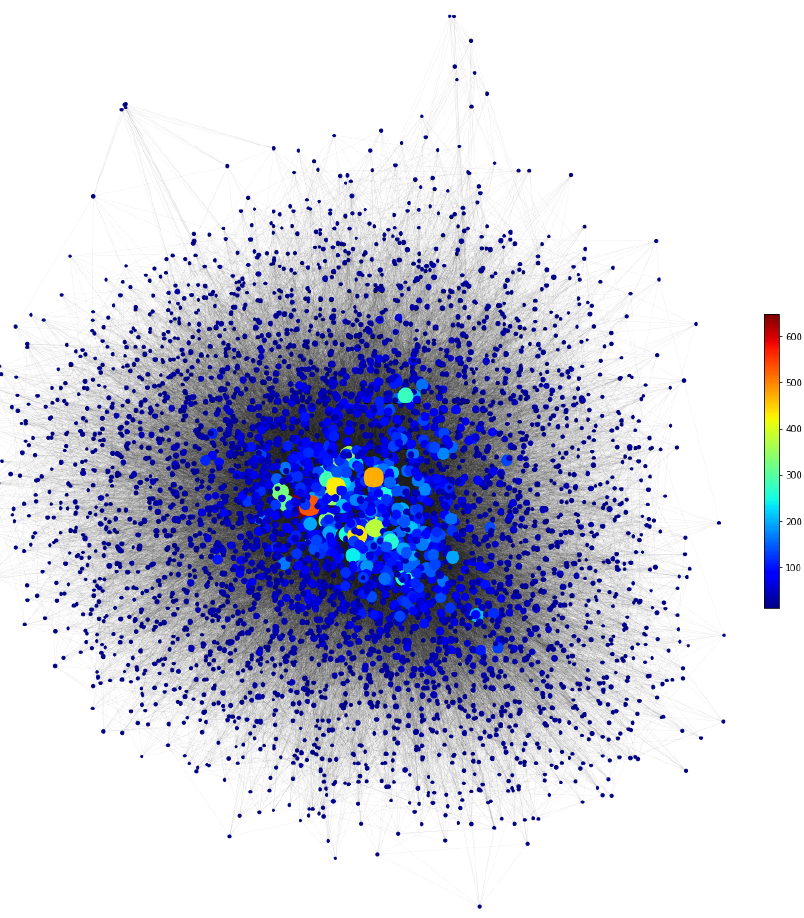}
\caption{Movie industry community graph}
\label{fig_4}
\end{figure*}

Following this approach, the entire graph has 50 million links for the existing dataset, so, due to computational limitations, the number of nodes was decreased using the Forest-fire based \cite{ForestFire} algorithm. In the first step, the one hundred most popular actors have been taken. Following this, titles where an actor has starred should be taken, then the rest of the cast, directors and casting directors should be taken, This should be repeated until one huge connected component is created. Finally, talent agents should be added to the graph. The generated graph has 59,443 nodes and 704,175 edges. The average clustering coefficient is 0.57, and the average shortest path is 8.

\subsection{Graph Description}

The graph (Figure \ref{fig_4}) has a core-periphery power law structure. The top three nodes by degree are casting directors: Mary Vernieu (casting director for “Deadpool” and another 421 titles), Kerry Barden (casting director for “The Spotlight”), Avy Kaufman (casting director for “The Sixth Sense”). The actor with the highest degree is Nicolas Cage.

\begin{figure*}
\centering
\begin{minipage}{.5\textwidth}
  \centering
  \includegraphics[width=1.0\linewidth]{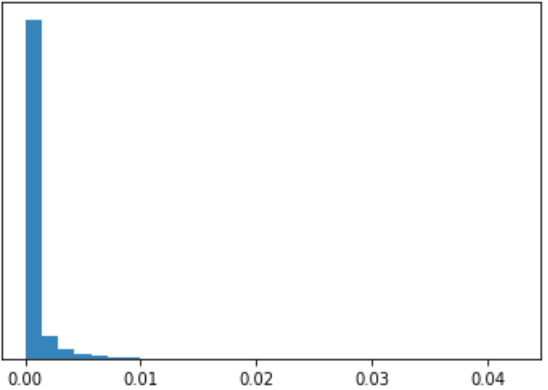}
  \caption{Degree centrality distribution}
  \label{fig_5}
\end{minipage}%
\begin{minipage}{.5\textwidth}
  \centering
  \includegraphics[width=1.0\linewidth]{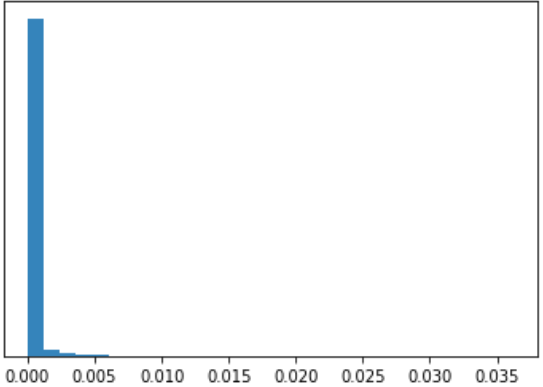}
  \caption{Betweenness centrality distribution}
  \label{fig_6}
\end{minipage}
\end{figure*}

Figure \ref{fig_5} shows that the graph follows the power law distribution of the node degrees - most of the nodes have small degree, which stand for new or unknown actors mostly. Most of the nodes have small eigenvalue centralities which can as nodes with high level of influence are not connected directly. Betweenness centrality distribution (Figure \ref{fig_6}) is concentrated in the area of small values as well, that is there is no nodes that strongly controls the network and the shortest paths often goes through different nodes.
Hubs and authorities algirithm shows that the top hubs are casting directors and the top authorities are highly popular actors, such as Nicole Kidman, Robert De Niro, Bruce Willice, Julianne Moore and Morgan Freeman. The results seem to be reasonable, because casting directors indeed work with a lot of people in movie industry and have a lot of contacts. 

\subsection{Random Walk Network Model}
Node2Vec is an algorithm that generates nodes representations in vector format. The algorithm creates low-dimensional nodes representations of a graph using random walks through a graph. The main idea of the method is that random walks through a graph may be considered as a sentence in a corpus in the following way \cite{node2vec}:
\begin{itemize}
    \item Every node is treated like a single word;
    \item Every walk through nodes of a graph is a sentence;
    \item Then, a skip-gram may be applied to every sentence, that was created by a random walk;
\end{itemize}

The main parameters of Node2Vec algorithm that may be controlled and adjusted are number of walks, length of each walk, p – the probability to return to the neighbour of the previous node, q – controls the probability to go away from the previous node during a random walk. In the current study we consider number of walks and its length as constants equal to 200 and 80 respectively, because it seems to be relevant for the graph of this size and computational power limits. Parameters $p$ and $q$ are considered from 1 to 4 each, so there are 16 models for each pair of parameters, since there is no intuitive understanding what probabilities will work in the best way for the current graph. Further in the research, each model will be used separately in prediction models. As a result, the model returns a 24-dimensional vector, that describes a node, which is a member of the movie industry community.

\section{Experiment}
First of all, the past average ratings of a movie title’s crew should be considered, because these variables were important in the previous studies reviewed in the chapter two. We will calculate average IMDb and RottenTomatoes rating of the movies for a director, casting director, writer and actors, so the they will be considered in the models and we will be able to see whether the additional SNA information about their position in the network will affect the results or not. Secondly, graph embeddings for movie team members should be obtained. In this study, we will consider vectors for directors, casting directors and four main actors of a movie. In case of boosting model the mean vector for actors will be taken and in case of neural network model embedding vectors will be taken as is. 

Also, because of reduction of the amount of data considered in chapter three the distribution of among buckets should be the same as in original dataset. In our case the original dataset has 53\% of the observations into “6-7” bucket, 37\% into “4-5” bucket, 8\% into “0-3” bucket and 2\% into “8-10” bucket. The reduced dataset has 56\% of the observations into “6-7” bucket, 35\% into “4-5” bucket, 7\% into “0-3” bucket and 2\% into “8-10” bucket, so the new distribution among classes is quite similar to the original.
 
Finally, since current dataset is imbalanced, we perform SMOTE algorithm \cite{SMOTE} in order to get equal number of samples in each class.

As for the comparison metric of the classification results the accuracy will be used \cite{accuracy}, because it is simple and suitable for the current multiclass classification problem.

\subsection{Random Forest Models}
The first model to test whether social network analysis features from movie industry community does improve the model is gradient boosting over random forest using CatBoost \cite{CatBoost} Python library.

The best results for the gradient boosting over random forest classification without SNA parameters were obtained with the following parameters:
\begin{itemize}
    \item Depth = 5;
    \item l2\_leaf\_reg = 1;
    \item Learning rate = 0.05
\end{itemize}
The best Accuracy = 0.68.

The best results for the gradient boosting over random forest classification with SNA parameters were obtained with the following parameters:

\begin{itemize}
    \item Depth = 4;
    \item l2\_leaf\_reg = 1;
    \item Learning rate = 0.03
\end{itemize}

The best Accuracy result is 0.74, it is 6\% gain due to social network features of the movie industry community.

\begin{table}
\centering
\begin{tabular}{|l|l|l|l|l|}
\hline \textbf{P,Q} & 1 & 2 & 3 & 4 \\ \hline
1 & 0.7404 & 0.7368 & 0.7363 & 0.7371  \\
2 & 0.7375 & \textbf{0.7410} & 0.7386 & 0.7373  \\
3 & 0.7407 & 0.7355 & 0.7371 & 0.7360 \\
4 & 0.7353 & 0.7386 & 0.7368 & 0.7393 \\
\hline
\end{tabular}
\caption{\label{table_node2vec} Accuracy results for different P and Q}
\end{table}

Comparing the results between models with different features produced by Node2Vec with different parameters, shows no much difference between the results, which is around 0.05 between the most extreme cases, however, the best was obtained with P = 2 and Q = 2 (Table \ref{table_node2vec}).

\begin{figure*}
\centering
\includegraphics[width=.85\linewidth]{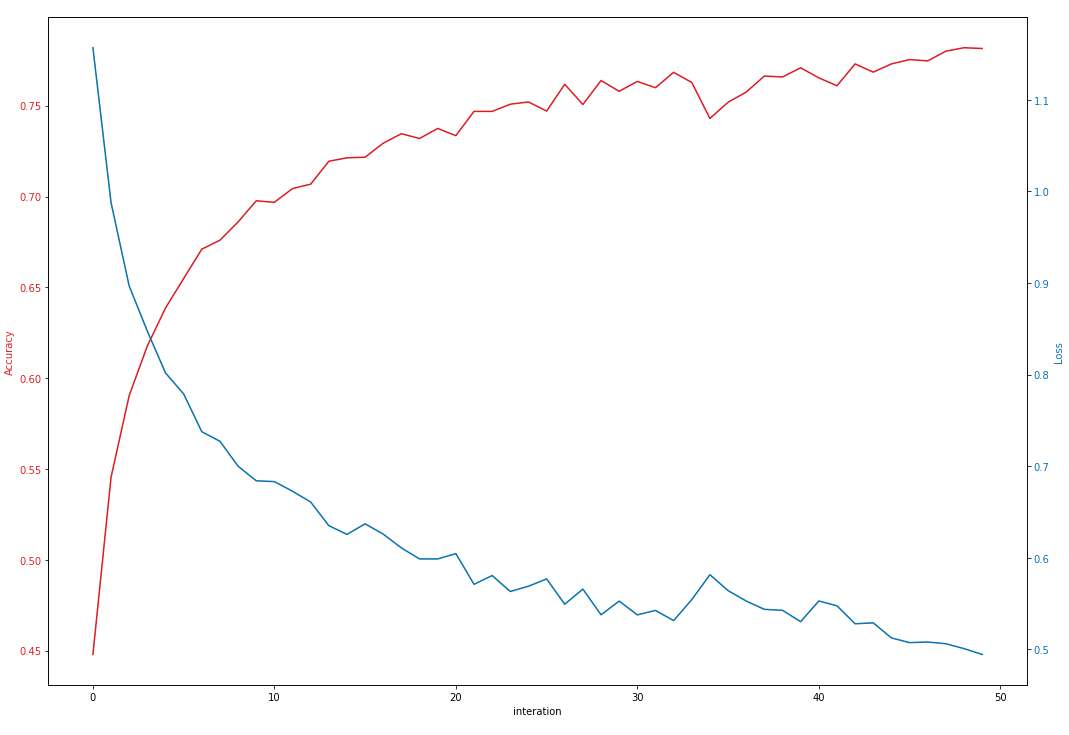}
\caption{Neural network model coverage with graph features}
\label{fig_7}
\end{figure*}

According to the feature importance, the most important network features were the ones describing directors’ nodes, the second were the ones describing casting directors’ and third were the ones describing actors’ nodes. Basically, that result shows the importance of casting director who is always in the background, because the importance of each feature indicates how much on average the prediction changes if the feature value changes.

\subsection{Random Forest Model}
The second test to check whether the SNA features affects the prediction results or not is to use random forest (RF) classification model. To find the optimal parameters grid search approach was used. 

The best results for the RF model without SNA parameters were achieved with the following set of the parameters:

\begin{itemize}
    \item Depth = 6;
    \item max\_features = auto;
    \item n\_estimators = 250;
\end{itemize}

The best results for the RF model with SNA parameters were achieved with the following set of the parameters:

\begin{itemize}
    \item Depth = 5;
    \item max\_features = auto;
    \item n\_estimators = 350;
\end{itemize}

The best accuracy result for the model without SNA features is 0.638 and the best accuracy results for the model with SNA features 0.671. So, as we can see the social network features improved the result but it is worse than the best accuracy score of the previous model.

\subsection{Decision Tree Model}

The third model is simple decision tree. Once again optimal parameters were chosen by grid search approach.

The best results for the model without SNA parameters were achieved with the following set of the parameters:

\begin{itemize}
    \item max\_depth = 300;
    \item max\_features = auto;
    \item criterion = gini;
\end{itemize}

The best results for the model with SNA parameters were achieved with the following set of the parameters:

\begin{itemize}
    \item Depth = 400;
    \item max\_features = auto;
    \item criterion = gini;
\end{itemize}

The best accuracy result without SNA is 0.611 and the best result with SNA is 0.601. So, in this case the results are quite similar, that is probably because of the high input dimension. 

\subsection{Neural Network Models}

The final test is to build neural network and compare the results considering different input vector for the actors, taking it as is. The classification model without movie industry community network features \cite{TensorFlow} architecture contains three layers with dropout:

\begin{table}
\centering
\begin{tabular}{|l|l|l|}
\hline Layer type & Shape & Param \\ \hline
Dense & 64 & 18688 \\
Dropout & 64 & 0 \\
Dense & 16 & 1040 \\
Dropout & 16 & 0 \\
Dense & 4 & 68 \\
\hline
\end{tabular}
\caption{\label{table_nn} Neural network architecture}
\end{table}

The result for the first model is 0.74, which is better comparing to the relative result obtained with the gradient boosting over random forest model

\begin{figure*}
\centering
\includegraphics[width=.85\linewidth]{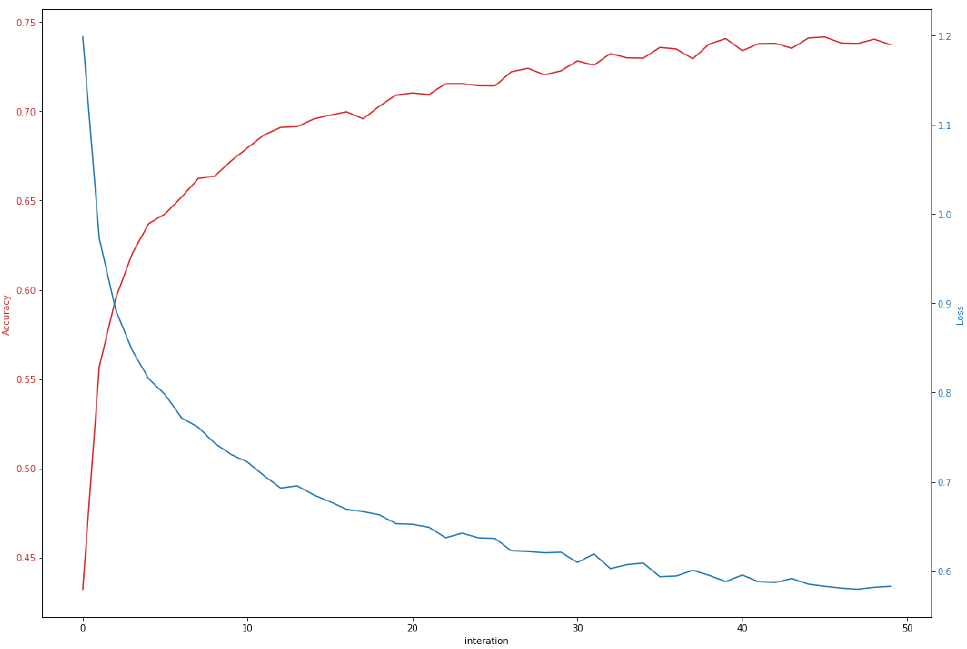}
\caption{Neural network model coverage with graph features}
\label{fig_8}
\end{figure*}

The same model with additional movie industry community network features has achieved result with Accuracy = 0.78 (Figure \ref{fig_8}), which is better than model without these features and better than boosting model created earlier. However, the difference between neural network models with and without graph features has become slightly smaller. 

\section{Conclusions}

Features extracted from movie industry community graph do improve predictions results as shown at Table \ref{table_results}. Director of a movie has more influence on movie success that casting director and actors. Also, we can notice that there is no much difference between Node2Vec model that have different random walk parameters.

\begin{table}
\centering
\begin{tabular}{|l|l|l|l|l|}
\hline & Without SNA features & With SNA features \\ \hline
RF + GB & 0.684 & 0.743 \\
RF & 0.638 & 0.671 \\
Decision Tree & 0.611 & 0.601 \\
NN & 0.741 & 0.781  \\
\hline
\end{tabular}
\caption{\label{table_results} Accuracy results for different models}
\end{table}

Also, as it can be seen in the \ref{table_results} most of the models showed that social network features obtained using Node2Vec to get the embeddings improved the prediction results even though data without SNA has information about movie crew - average rating for the previous titles, so it can be concluded that SNA features are significant and a position in the movie industry community does affect future movie success.

In the present research the movie industry community have been studied by actor-casting director-talent agent-direct graph analysis and the following goals have been completed:
\begin{itemize}
    \item The unique data about casting directors and talent agents has been collected by scrapping web pages;
    \item The network structure of the movie community has been presented in terms of graph and analyzed providing information that describes how this community works;
    \item The embeddings that describe nodes have been obtained;
    \item The models predicting IMDb score has been created and the information of the different combinations of the features sets has been collected and analyzed.
\end{itemize}

Based on the obtained results by achieving these goals it may be concluded that movie community industry does follow the common social network properties, such as power law distribution, that is there are small number of members who has a significant number of connections with other people from the community. Also, it has been shown that casting directors are important member of the movie community as they are network hubs. Moreover, during the current research is has been shown that features that describes network structure of the movie industry community does influence the IMDb rating. The most important is the group of features regarding director of a movie titles, the second most important is the group of features that contains information about casting directing directors, so, these people do affect the movie success. Also, as it was described in the previous studies, the network model shows better results predicting IMDb score.

However, this study has some limitations, such as reduced dataset due to computational resources, so the future researches may include:

\begin{itemize}
    \item Analysis of the whole graph with all principals of a movie, because it may give more insights;
    \item Usage of the social network model for best strategies in order to get the cast, because it may be more efficient to spend resources on a popular casting director or actor or talent agent that will bring other people because they are valuable members of the community
\end{itemize}

\section{Acknowledgements}
The article was prepared within the framework of the HSE University Basic Research Program.

\bibliographystyle{unsrt}
\bibliography{custom}

\end{document}